\documentclass[amsmath,amssymb,twocolumn,prl,showpacs,nofootinbib,floatfix,superscriptaddress]{revtex4-2}
\usepackage{float}
\usepackage{soul}
\usepackage{amsmath}
\usepackage{graphicx}% Include figure files
\usepackage{dcolumn}% Align table columns on decimal point
\usepackage{bm}% bold math
\usepackage{natbib}
\usepackage{tikz}
\usepackage{color}
\usepackage{subfigure}
\usepackage{longtable}
\usepackage{dcolumn}
\usepackage{etoolbox}
\usepackage{lmodern}
\usepackage{slantsc}
\usepackage{textcomp}
\usepackage[colorlinks, linkcolor={blue},citecolor={blue},breaklinks]{hyperref}% add hypertext capabilities
\begin{document}

\preprint{APS/123-QED}

\title{Constraining the $^{30}$P($p,\gamma)^{31}$S reaction rate in ONe novae via the weak, low-energy, $\beta$-delayed proton decay of $^{31}$Cl}% Force line breaks with \\

\author{T. Budner}
\email{budner@nscl.msu.edu}
\affiliation{National Superconducting Cyclotron Laboratory, Michigan State University, East Lansing, MI 48824, USA}
\affiliation{Department of Physics and Astronomy, Michigan State University, East Lansing, MI 48824, USA}

\author{M. Friedman}
%\email{moshe.friedman@mail.huji.ac.il}
\affiliation{National Superconducting Cyclotron Laboratory, Michigan State University, East Lansing, MI 48824, USA}
\affiliation{Racah Institute of Physics, Hebrew University, Jerusalem, Israel 91904}

\author{C. Wrede}
\email{wrede@nscl.msu.edu}
\affiliation{National Superconducting Cyclotron Laboratory, Michigan State University, East Lansing, MI 48824, USA}
\affiliation{Department of Physics and Astronomy, Michigan State University, East Lansing, MI 48824, USA}

\author{B. A. Brown}
\affiliation{National Superconducting Cyclotron Laboratory, Michigan State University, East Lansing, MI 48824, USA}
\affiliation{Department of Physics and Astronomy, Michigan State University, East Lansing, MI 48824, USA}

\author{J. Jos\'e}
\affiliation{Departament de F\'isica, Universitat Polit\`ecnica de Catalunya, E-08019 Barcelona, Spain}
\affiliation{Institut d'Estudis Espacials de Catalunya,  Universitat Polit\`ecnica de Catalunya, E-08034 Barcelona, Spain}

\author{D. P\'erez-Loureiro}
\affiliation{National Superconducting Cyclotron Laboratory, Michigan State University, East Lansing, MI 48824, USA}

\author{L. J. Sun}
\affiliation{National Superconducting Cyclotron Laboratory, Michigan State University, East Lansing, MI 48824, USA}
\affiliation{School of Physics and Astronomy, Shanghai Jiao Tong University, Shanghai 200240, China}
\author{J. Surbrook}
\affiliation{National Superconducting Cyclotron Laboratory, Michigan State University, East Lansing, MI 48824, USA}
\affiliation{Department of Physics and Astronomy, Michigan State University, East Lansing, MI 48824, USA}

\author{Y. Ayyad}
\affiliation{National Superconducting Cyclotron Laboratory, Michigan State University, East Lansing, MI 48824, USA}
\affiliation{IGFAE, Universidade de Santiago de Compostela, E-15782 Santiago de Compostela, Spain}

\author{D. W. Bardayan}
\affiliation{Department of Physics, University of Notre Dame, Notre Dame, IN 46556, USA}

\author{K. Chae}
\affiliation{Department of Physics, Sungkyunkwan University, Seoul, South Korea}

\author{A. A. Chen}
\affiliation{Department of Physics and Astronomy, McMaster University, Hamilton, ON L8S 4L8, Canada}

\author{K. A. Chipps}
\affiliation{Physics Division, Oak Ridge National Laboratory, Oak Ridge, TN 37830-37831, USA}
\affiliation{Department of Physics and Astronomy, University of Tennessee, Knoxville, TN 37996, USA}

\author{M. Cortesi}
\affiliation{National Superconducting Cyclotron Laboratory, Michigan State University, East Lansing, MI 48824, USA}

\author{B. Glassman}
\affiliation{National Superconducting Cyclotron Laboratory, Michigan State University, East Lansing, MI 48824, USA}
\affiliation{Department of Physics and Astronomy, Michigan State University, East Lansing, MI 48824, USA}

\author{M. R. Hall}
\affiliation{Department of Physics, University of Notre Dame, Notre Dame, IN 46556, USA}

\author{M. Janasik}
\affiliation{National Superconducting Cyclotron Laboratory, Michigan State University, East Lansing, MI 48824, USA}
\affiliation{Department of Physics and Astronomy, Michigan State University, East Lansing, MI 48824, USA}

\author{J. Liang}
\affiliation{Department of Physics and Astronomy, McMaster University, Hamilton, ON L8S 4L8, Canada}

\author{P. O'Malley}
\affiliation{Department of Physics, University of Notre Dame, Notre Dame, IN 46556, USA}

\author{E. Pollacco}
\affiliation{D\'epartement de Physique Nucl\'eaire, IRFU, CEA, Universit\'e Paris-Saclay, F-91191, Gif-sur-Yvette, France}

\author{A. Psaltis}
\affiliation{Department of Physics and Astronomy, McMaster University, Hamilton, ON L8S 4L8, Canada}

\author{J. Stomps}
\affiliation{National Superconducting Cyclotron Laboratory, Michigan State University, East Lansing, MI 48824, USA}
\affiliation{Department of Physics and Astronomy, Michigan State University, East Lansing, MI 48824, USA}

\author{T. Wheeler}
\affiliation{National Superconducting Cyclotron Laboratory, Michigan State University, East Lansing, MI 48824, USA}
\affiliation{Department of Physics and Astronomy, Michigan State University, East Lansing, MI 48824, USA}

\date{\today}

\begin{abstract}
The $^{30}$P$(p,\gamma)^{31}$S reaction plays an important role in understanding nucleosynthesis of $A\geq 30$ nuclides in oxygen-neon novae. The Gaseous Detector with Germanium Tagging was used to measure $^{31}$Cl $\beta$-delayed proton decay through the key $J^{\pi}=3/2^{+}$, 260-keV resonance. The intensity $I^{260}_{\beta p} = 8.3^{+1.2}_{-0.9} \times 10^{-6}$ represents the weakest $\beta$-delayed, charged-particle emission ever measured below 400 keV, resulting in a proton branching ratio of $\Gamma_p / \Gamma = 2.5^{+0.4}_{-0.3} \times 10^{-4}$. By combining this measurement with shell-model calculations for $\Gamma_{\gamma}$ and past work on other resonances, the total $^{30}$P$(p,\gamma)^{31}$S rate has been determined with reduced uncertainty. The new rate has been used in hydrodynamic simulations to model the composition of nova ejecta, leading to a concrete prediction of $^{30}$Si/$^{28}$Si excesses in presolar nova grains and the calibration of nuclear thermometers.
\end{abstract}

\maketitle

Classical novae occur in stellar binary systems involving a compact white dwarf (WD) that siphons hydrogen-rich material from its companion star to form an accretion disk. The accreted mass is heated, compressed, and mixed with the outer layers of the underlying WD until it eventually ignites in a thermonuclear runaway \cite{Chomiuk2021}. These explosive events eject freshly synthesized nuclei into the interstellar medium, contributing to the chemical evolution of the Galaxy \cite{Gehrz1998}. Novae are good test cases for models of explosive nucleosynthesis since they occur frequently in the Milky Way, with about a dozen observed annually \cite{NASA2021}. In addition, the path of nucleosynthesis in novae is close enough to stability that most of the relevant thermonuclear reactions rates can be determined experimentally. Sensitivity studies suggest that $^{30}$P($p$,$\gamma$)$^{31}$S is the dominant nuclear physics uncertainty impacting the production of mass number $A \geq 30$ nuclides in oxygen-neon (ONe) classical novae \cite{Jose2001, Jose2006}. As a result, this particular reaction rate affects the identification of certain presolar grains, nuclear mixing meters, and the calibration of nova thermometers. 

The $^{30}$P($p$,$\gamma$)$^{31}$S reaction serves as a bottleneck for the production of heavier elements in a complex reaction network of competing proton captures and $\beta^+$ decays \cite{Jose2016}.
The 2.5-minute half-life of $^{30}$P, which decays to stable $^{30}$Si \cite{Wilson1980}, is on the same order as the timescales of the thermonuclear runaway. Thus, a relatively slow reaction rate would predict huge $^{30}$Si excesses in nova ejecta, while a faster rate could lead to the synthesis of intermediate-mass nuclides up to $A \approx 40$ as well as more modest $^{30}$Si/$^{28}$Si ratios \cite{Jose2004}. This might explain several presolar grains found in the Murchison carbonaceous meteorite, whose anomalous isotopic signatures cannot be definitively attributed to a known stellar source. These grains are characterized by reduced $^{12}$C/$^{13}$C and very low $^{14}$N/$^{15}$N ratios as well as large enhancements in $^{30}$Si/$^{28}$Si ratios when compared to typical solar abundances, leading some to hypothesize that these grains condensed in the ejecta from ONe novae \cite{Amari2001}. However, large uncertainties in the $^{30}$P($p$,$\gamma$)$^{31}$S rate have prevented a concrete prediction of $^{30}$Si abundances for theoretical nova grains.

The amount of Si produced in novae can also inform the mixing mechanisms of astrophysical models. How exactly fuel from the donor star mixes with the dense material on the surface of the WD is unclear \cite{Casanova2011,Casanova2016,Casanova2018}, but the extent of the mixing influences the predicted nucleosynthetic yields. Different mixing fractions result in a range of chemical abundances, and constraining the $^{30}$P($p$,$\gamma$)$^{31}$S rate reduces uncertainties in the expected Si/H ratios of nova ejecta \cite{Kelly2013}. Similarly, elemental abundances observed via ultraviolet, optical, and infrared spectroscopy can be used to constrain peak nova temperatures. Specifically, the ratios O/S, S/Al, O/P, and P/Al are good candidates for thermometers, as they exhibit steep, monotonic dependences on temperature. The crucial $^{30}$P($p$,$\gamma$)$^{31}$S reaction rate remains the dominant nuclear uncertainty limiting their accuracy and precision \cite{Downen2013}.

In lieu of a rate based on measured resonance properties, this reaction is approximated using the Hauser-Feshbach (HF) statistical model \cite{Hauser1952}. However, this method is not expected to be accurate for $^{30}$P($p$,$\gamma$)$^{31}$S across peak nova temperatures ($T_{\text{peak}} = 0.1-0.4$ GK) \cite{Iliadis2001}, and astrophysical studies will often vary this rate by orders of magnitude in simulation to account for its large uncertainty. The HF method assumes the nuclear level density is
sufficiently high such that it can be modeled as a continuum, but for many cases, especially near shell closures and the drip lines, radiative captures into narrow, isolated resonances must be considered individually \cite{Rauscher2000}. In the case of $^{30}$P($p$,$\gamma$)$^{31}$S, the rate is dominated by proton capture on the ground state of $^{30}$P into low-lying resonances $\lesssim$ 600 keV above the $^{31}$S proton-emission threshold ($S_p$ = 6130.65(24) keV) \cite{Wang2021}, as novae are not hot enough to appreciably populate excited states in $^{30}$P nuclei. One needs to know the strengths of individual resonances within the Gamow window to determine the total thermonuclear rate \cite{Iliadis2007}.

Currently, $^{30}$P beams cannot be produced with the intensities needed to measure this reaction directly at the astrophysically relevant low energies. Over the past two decades, significant theoretical \cite{Brown2014} and experimental effort has been devoted to studying the level structure of $^{31}$S in an effort to determine resonance properties. Single-nucleon transfer reactions \cite{Vernotte1999, Ma2007, Wrede2009, Irvine2013, Parikh2016, Kankainen2017, Burcher2019, Setoodehnia2020}, in-beam $\gamma$-ray spectroscopy \cite{Jenkins2005, Jenkins2006, Vedova2007, Pattabiraman2008, Tonev2011, Doherty2012, Doherty2014}, $\beta$-decay measurements \cite{Kankainen2006, Saastamoinen2011, Kankainen2014, Saastamoinen2016}, and the charge-exchange reaction $^{31}$P($^3$He,$t$)$^{31}$S \cite{Wrede2007,Wrede2009,Parikh2011} have all been employed to constrain the spins, parities, and decay widths of $^{31}$S excited states. It is likely that all of the potentially important resonances contributing to $^{30}$P($p$,$\gamma$)$^{31}$S in novae have been observed using various nuclear spectroscopy techniques \cite{Wrede2014}. 

In a $^{31}$Cl $\beta$-delayed $\gamma$ decay experiment performed at the National Superconducting Cyclotron Laboratory (NSCL), a $J^{\pi}=3/2^+$ level with an excitation energy $E_x = 6390.2(7)$ keV was discovered in the $^{31}$S compound nucleus \cite{Bennett2016}. The $J^{\pi}=1^+$ ground state of the $^{30}$P target nucleus \cite{Hafner1974} makes this an $\ell=0$ resonance for $^{30}$P($p$,$\gamma$)$^{31}$S. Thus, proton capture to this state is not inhibited by a centrifugal barrier. Furthermore, its 260-keV resonance energy is in the middle of the Gamow window for peak nova temperatures, suggesting this resonance could dominate the total thermonuclear rate. To experimentally determine the strength of this resonance, both the lifetime and the proton branching ratio of this state need to be measured. In this Letter, we present the results of a radioactive beam experiment conducted at NSCL to measure the critical resonance's proton branching ratio $\Gamma_p / \Gamma$. The 6390-keV excited state in $^{31}$S with isospin $T=1/2$ is strongly populated by $^{31}$Cl $\beta^+$ decay due to isospin mixing with the isobaric analog state (IAS) of the $^{31}$Cl ground state $(T=3/2)$ \cite{Bennett2016}. However, the intensity of the $\beta$-delayed proton decay is suppressed by the Coulomb barrier at such low energies, making this weak proton branch challenging to measure. 

The Gaseous Detector with Germanium Tagging (GADGET) was developed specifically to measure this $\beta$-delayed proton decay but has already been used for other cases \cite{Friedman2019,Friedman2020,Sun2021}. GADGET utilizes a customized cylindrical, gas-filled proportional counter called the Proton Detector (PD). Similar in concept to the AstroBox instrument \cite{Pollacco2013}, this detector was developed to mitigate the substantial low-energy backgrounds and summing effects encountered from the interactions between $\beta^+$ particles and solid-state Si detectors \cite{Saastamoinen2011}. Equipped with 13 charge-sensitive, Micromegas detection pads, the inner five pads measure the intensities of charged-particle decays that deposit their full energy in the active region, a cylindrical volume spanning the length of the gaseous chamber, whose 40-mm radius corresponds to the boundary between the active, inner pads and outer, veto pads. The eight outer pads are used to veto higher-energy protons, whose energy deposition outside the active region exceeds the trigger threshold. The PD is surrounded by the Segmented Germanium Array (SeGA), which consists of 16 high-purity Ge crystals arranged into two rings of eight individual $\gamma$ detectors \cite{Mueller2001}. GADGET couples these detection systems, enabling proton-$\gamma$ coincidence analysis.

The Coupled Cyclotron Facility accelerated a 75-pnA primary beam of $^{36}$Ar to 150 MeV/u, impinging it on a 1645-mg/cm$^2$ thick Be production target. The A1900 fragment separator was used to purify the secondary beam via magnetic rigidity separation \cite{Morrissey2003}. The Radio Frequency Fragment Separator (RFFS) further purified the beam, resulting in a 65\% pure $^{31}$Cl beam upon exiting the RFFS at 6400 pps \cite{Bazin2009}. The main contaminants, in decreasing order of intensity, were $^{28}$Si, $^{30}$S, and $^{29}$P, but none of these are $\beta$-delayed particle emitters. Intermittently between measurement runs, a single Si PIN detector was used to determine the beam energy deposition a meter upstream of GADGET. We used version 1.2 of the \textsc{atima} program in \textsc{lise++} to calculate the beam energy loss in 300 $\mu$m of Si and compared this to the observed energy loss in the PIN detector, confirming $^{31}$Cl and stable $^{28}$Si as the two main species delivered to the setup \cite{Tarasov2008}. Located directly in front of the PD, a 0.75-mm thick Al beam-energy degrader was manually rotated to an angle that optimized the longitudinal implantation distribution of $^{31}$Cl in the center of GADGET. The beam entered the PD chamber through a thin, Kapton window and was implanted in the 808-Torr gas mixture of 90\% Ar and 10\% CH$_4$ (P10). The circular entrance aperture has a 25.4-mm radius and is aligned with the center of the Micromegas pad plane. Beam particles thermalized after entering the PD diffuse radially under Brownian motion until they decay.

\begin{figure}
	\centering
	\includegraphics[width=0.47\textwidth]{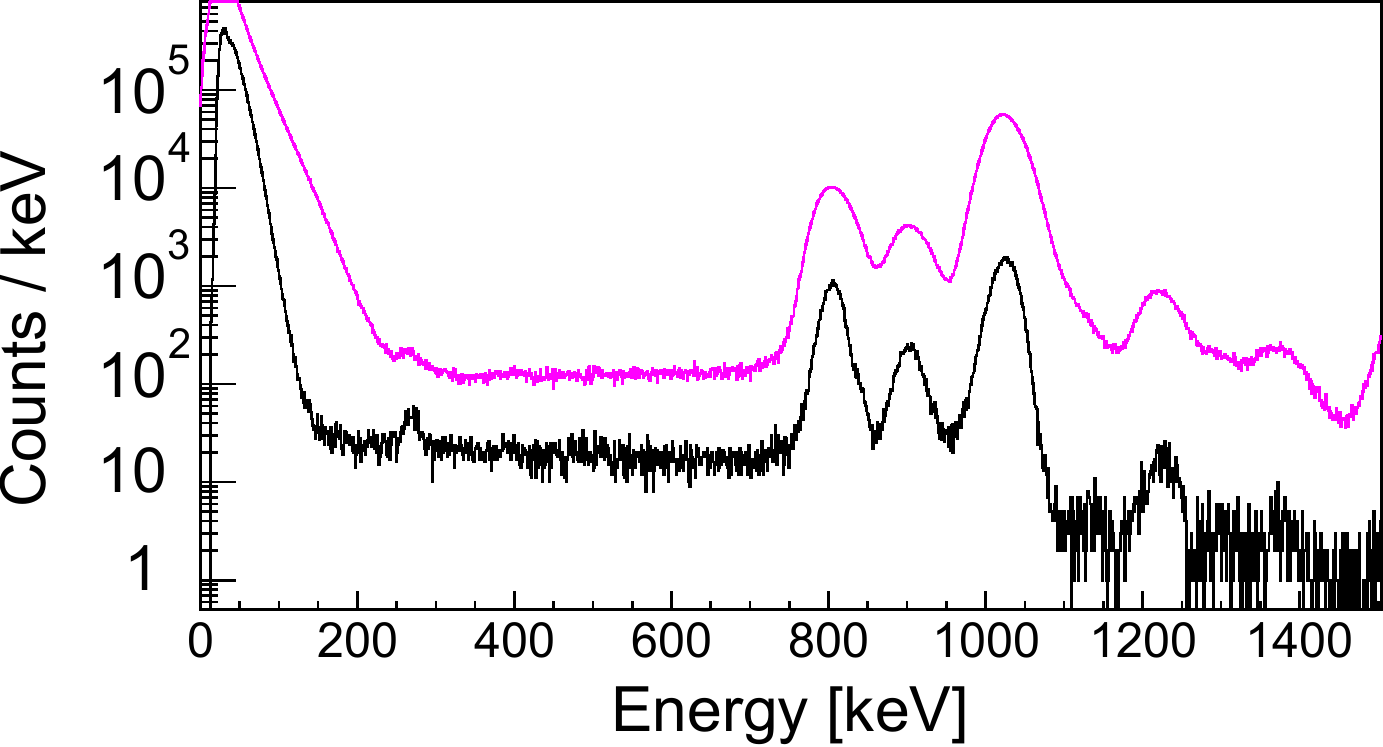}
	\caption{[Color online] $^{31}$Cl $\beta$-delayed proton spectrum measured by only the central detector pad [black] and for event-level summing of the five inner detector pads [grey (pink online)] up to 1.5 MeV. The energy spectrum sums the ionization deposited in the P10 gas from both the decay protons and recoiling $^{30}$P nuclei. $\beta^+$ particles are responsible for the large background at low energies and can also sum with ionization produced by proton tracks, leading to a detector response that is skewed to the right; this effect is larger in the combined-pad spectrum due to the effective increase in detection pad area.}
	\label{fig:h0h6}
\end{figure}

The dataset referenced in this Letter consists of over 86 hours of accumulated beam time. Events from groups of hour-long runs were added together to improve statistics before fitting the two largest peaks in the proton spectra and applying a linear gain-matching procedure. Fig. \ref{fig:h0h6} shows the $\beta$-delayed proton spectra for center-of-mass decay energies. For our energy calibration, we used the resonance energies of the three strongest $\beta$-delayed proton decays in the spectrum: 806, 906, and 1026 keV. These energies are taken from evaluated nuclear data tables \cite{Batchelder2020}, which adopt values from Ref. \cite{Saastamoinen2011}, each with a 2-keV uncertainty. Similarly for SeGA, 15 of the 16 detectors were calibrated individually, relative to known $\gamma$ rays from room background radiation; a single SeGA detector exhibited poor energy resolution and was excluded from this analysis. Gain-matching was performed on a run-by-run basis to produce a cumulative $\gamma$ spectrum.

We observe $\beta$-delayed protons emitted from the 6390-keV level in $^{31}$S for the first time, as shown in Fig. \ref{fig:h0h6}. Using the $\gamma$-tagging capabilities of GADGET, we conclude that these proton decays are not in coincidence with $\gamma$ transitions, confirming that this proton emission populates the $^{30}$P ground state. Fitting only the central pad spectrum in Fig. \ref{fig:h0h6}, which has a reduced $\beta$ background, sharper peaks, and thus a more precise energy calibration, we measure this decay energy to be $E_r = 273(10)$ keV, consistent to within 1.4 standard deviations of the 260-keV resonance energy measurement by Bennett \textit{et al.} \cite{Bennett2016}. The largest source of uncertainty in energy is 8.5 keV, which results from extrapolating the linear calibration function. This is added in quadrature with the 5-keV systematic error associated with the pulse height defect \cite{Ziegler2010} and with the 1-keV statistical uncertainty of the fit.

We model the event-level, combined-pad PD response function as an exponentially modified Gaussian distribution with a high-energy tail to account for the effect of $\beta$ summing. We tested this model on proton peaks $>$700 keV, parameterizing the shape of this distribution and its dependence on decay energy. Then, we used this function to fit the 260-keV proton peak in Fig. \ref{fig:h0h6}, modeling the low-energy $\beta$ background as exponential and the relatively flat background $>$300 keV as linear. We determined the total number of low-energy, $\beta$-delayed protons observed in this experiment to be $N_{\beta p}^{260} = 2731(203)$. For the purposes of normalization, we adopt the recommended literature value $I_{\beta p}^{1026} = 0.0131(2)$ for the strongest $\beta$-delayed proton decay peak in our spectrum at 1 MeV \cite{Saastamoinen2011,Batchelder2020}. To determine the $\beta$-delayed proton decay intensity of the 260-keV resonance, we use the relation $I_{\beta p}^{260} = (N_{\beta p}^{260} / N_{\beta p}^{1026}) \times (\epsilon_p^{1026} / \epsilon_p^{260}) \times I_{\beta p}^{1026}$, where $\epsilon_p^{1026} / \epsilon_p^{260}$ is the ratio of PD efficiencies at the notated resonance energies. The cumulative proton spectrum was fit over $350-1100$ keV to determine the number of 1-MeV protons $N_{\beta p}^{1026}=3.16(2) \times 10^6$, modeling the three large calibration peaks as exponentially modified Gaussian distributions on top of a linear background. 

To evaluate the efficiency of GADGET, we developed a geometric, Monte Carlo simulation to calculate the probability of detecting an event as a function of proton energy. We simulated ionization tracks for $10^4$ isotropic decays at each proton energy to achieve a statistical uncertainty of 1\% in $\epsilon_p$. Stopping powers for $^1$H ions in 808-Torr, P10 gas were calculated using \textsc{srim}, which quotes a 4\% uncertainty \cite{Ziegler2010}. This translates into a 1\% systematic uncertainty on the lower limit of $\epsilon_p^{1026} / \epsilon_p^{260}$ and a 2\% uncertainty on the upper limit. The initial transverse positions of these simulated proton emissions were randomly sampled from a 2D Gaussian beam spot, whose centroid and width parameters were deduced from a $\chi^2$-minimization procedure using the relative number of measured proton counts on each PD pad as input. The uncertainty associated with each of these parameters is about $\pm 6$ mm and leads to a 2\% lower limit error and a 5\% upper limit error. The systematic uncertainty associated with the diffusion of beam particles is also about the same size. The upper limit on detection efficiency assumes no beam diffusion, effectively constraining the beam radius to the size of the entrance window. Allowing the $^{31}$Cl distribution to diffuse transversely beyond the aperture radius for two half-lives represents the lower limit of the efficiency, which corresponds to a radial displacement of $\approx$ 5 mm.

The last source of systematic uncertainty in the Monte Carlo efficiency model is related to the veto trigger threshold. Simulated proton events whose full energy is confined to the active region of the detector are recorded as measured events, but if a proton track ionizes too many electrons outside the active region, the proton event is vetoed. For each simulated event, a multitude of electrons, proportional to the proton energy, are randomly generated along the length of the ionization track. Each electron position is evaluated to determine the proton event's veto status. This is complicated by the fact that ionization electrons spread out transversely in time according to the relation $\sigma = \sqrt{4Dt}$ \cite{Sauli2014}, where $t$ for each simulated decay event is randomly sampled from the longitudinal beam distribution as measured in ionization drift time; the electron diffusion coefficient $D=9116(273)$ cm$^2$/s was calculated using \textsc{magboltz} \cite{Biagi1999}. Multiple simulations were performed for a range of veto conditions, spanning the minimum and maximum gain-matched energy thresholds (5-20 keV) across all PD pads over the entire experiment. We estimate a systematic uncertainty of 3-4\% in both directions for the relative detection efficiency as a result of the veto threshold.

Combining the final efficiency ratio $\epsilon_p^{1026} / \epsilon_p^{260} = 0.73^{+0.09}_{-0.05}$ with measured proton counts and the adopted literature intensity for the 1-MeV protons, we arrive at $I_{\beta p}^{260} = 8.3 ^{+1.2}_{-0.9} \times 10^{-6}$, the weakest $\beta$-delayed proton intensity ever measured for resonances below 400 keV. Such low-lying, proton-unbound states are typically dominated by $\gamma$ decay, and a previous measurement of $^{31}$Cl $\beta$ decay determined the intensity of $\beta$-delayed $\gamma$ emission through the 6390-keV state to be $I_{\beta \gamma}^{6390} = 0.0338(18)$ \cite{Bennett2018}. In the limit where $\Gamma_{\gamma} \gg \Gamma_p$, we can compute the proton branching ratio as $\Gamma_p / \Gamma \approx I_{\beta p}^{260} / I_{\beta \gamma}^{6390} = 2.5^{+0.4}_{-0.3} \times 10^{-4}$. 

Without a finite lifetime measurement, we evaluate $\Gamma_{\gamma}$ theoretically in order to calculate the resonance strength. For the USDC Hamiltonian, the strongest isospin mixing with the IAS of the $^{31}$Cl ground state comes from a $T=1/2$ level in $^{31}$S, which theory predicts to be 300 keV below the $T=3/2$ IAS. The isospin-mixing matrix element for these levels $V_{\text{theory}}=36$ keV is in good agreement with the experimental value $V_{\text{exp}}=41(1)$ keV \cite{Magilligan2020}. Theory predicts $\gamma$-decay widths of 190 meV for the $T=1/2$ state and 920 meV for the $T=3/2$ state. However, the observed $T=1/2$ state, corresponding to the 6390-keV level in $^{31}$S, lies above the $T=3/2$ IAS. The mixing of these two states depends on the energy difference between them, which is determined by the strong interaction. We added a term in the Hamiltonian proportional to the $T^{2}$ operator to move the $T=1/2$ state up by 410 keV. After this shift, the new partial widths are $\Gamma_{\gamma}^{6390} = 490$ meV and $\Gamma_{\gamma}^{\textrm{IAS}} = 430$ meV. The sum of the widths is not exactly the same due to some interference with other $T=1/2$ states that do not have a strong isospin mixing. The uncertainty in $\Gamma_{ \gamma }$ is about 50 meV, which comes from the four different Hamiltonians derived in Ref. \cite{Magilligan2020}. The $\gamma$ decay of the $T=1/2$ state is dominated by a 66\% branch to the lowest $J^{\pi}=5/2^+$ state with $B(M1)=0.48 \mu _{N}^{2}$. The $M1$ decay matrix element is $ M = \sqrt{(2J_{i}+1) B(M1)}= 1.38 \mu_{N}$. The error in $M$, coming from a comparison of other experimental values of $M$, is about 0.4; see Fig. 4 in Ref. \cite{Richter2008}. This leads to an estimated uncertainty of 280 meV. Thus, we adopt $\Gamma_{\gamma} = 490(280)$ meV for the $T=1/2$ resonance state. 

\begin{figure}
	\centering
	\includegraphics[width=0.46\textwidth]{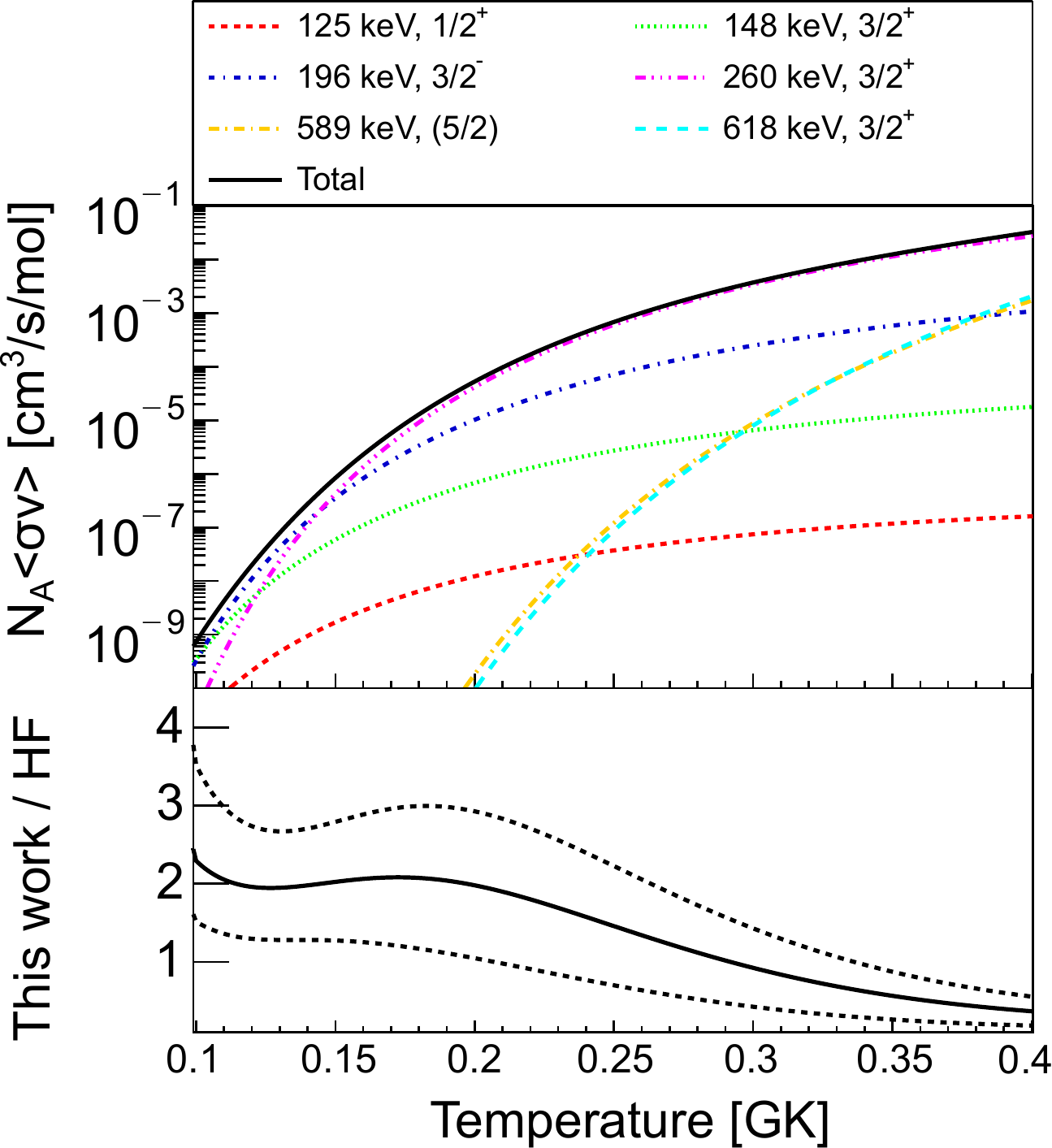}
	\caption{[Color online] (a) Contributions of individual resonances to the $^{30}$P$(p,\gamma)^{31}$S reaction rate and the total summed thermonuclear rate [solid black] plotted over peak nova temperatures. (b) The ratio between the experimental resonant reaction rate and the Hauser-Feshbach statistical rate \cite{Rauscher2000}. The solid curve represents the recommended central rate, while the dashed curves indicate the upper and lower limits on the resonant rate calculation.}
	\label{fig:rates}
\end{figure}

Using this value and the measured branching ratio, we compute a resonance strength of $\omega\gamma = 80(48)$ $\mu$eV. Combining this result with experimental and theoretical information for several other $^{31}$S levels, we calculated the total $^{30}$P$(p,\gamma)^{31}$S rate. A summary of the most recent experimental constraints on relevant resonances is provided by Kankainen $\textit{et al.}$, reporting spectroscopic factors for several states in the Gamow window \cite{Kankainen2017}, which we adopt. For potentially significant resonances lacking experimental data, we appeal to theoretical strength calculations \cite{Brown2014}. The contributions of the most significant resonances to the total rate are shown in Fig. \ref{fig:rates}. The 260-keV, $3/2^+$ resonance measured in this work dominates the total rate over most of the peak ONe nova temperature range.

We utilize the 1D, fully hydrodynamic code \textsc{shiva} to simulate a series of nova explosions involving a 1.35-\(M_\odot\) ONe WD \cite{Jose2016,Jose1998}. These calculations were performed using our recommended reaction rate and the $1\sigma$ upper and lower rate limits to quantify the nuclear uncertainties in the nova model. We express our predicted Si isotopic ratio for ONe nova ejecta in permil deviation from solar abundances: $\delta(^{30}$Si/$^{28}$Si) = $+1.14^{+0.93}_{-0.35} \times 10^4$\textperthousand. We are in agreement with previous simulations using the HF statistical rate but with substantially reduced uncertainties. Previous predictions of $^{30}$Si excesses varied by a factor of $\approx 6$ between the lower limit and the nominal rate, while the uppermost limits of the HF rate even predicted deficits \cite{Jose2004}. Thus, we can conclusively say for the first time that ONe novae should produce $^{30}$Si excesses in their ejecta for the heaviest WD masses.

Finally, we are able to reduce uncertainties in predicted abundances used to calibrate nova thermometers. Based on observed N/O and O/S ratios, the nova V838 Herculis is reported to reach temperatures $T_{\text{peak}} = 0.30-0.31$ GK, corresponding to a WD mass of $M_{\text{WD}} = 1.34-1.35$ \(M_\odot\) \cite{Downen2013}. We simulated an ONe nova explosion for the same WD mass and found abundance (mass fraction) ratios O/S = $0.86^{+0.12}_{-0.04}$ and S/Al = $15.7^{+0.9}_{-2.2}$ to agree with astronomical observation to within $1\sigma$. For a 0.31-GK peak temperature, the variation, or quotient of upper and lower limits, in predicted mass fraction ratios caused by varying the $^{30}$P$(p,\gamma)^{31}$S rate within its reported error bars has been reduced by factors of about $2-4$ for O/S, S/Al, O/P, and P/Al ratios. The main limitation in the accuracy of these thermometers is now the typical precision of abundance observations.

The present work represents both a technical achievement for measuring such a weak, low-energy, $\beta$-delayed proton decay, as well as a significant reduction in nuclear uncertainties for modeling nova nucleosynthesis. The dominant source of uncertainty in the recommended $^{30}$P$(p,\gamma)^{31}$S rate is now the theoretical $\Gamma_{\gamma}$ value for the 6390-keV level in $^{31}$S, motivating experiments to determine this state's lifetime, since $\Gamma_{\gamma} \approx \hbar/\tau.$ Nevertheless, we are able to demonstrate production of $^{30}$Si excesses in state-of-the-art simulations for the most energetic ONe nova explosions, and we present a new calibration point for nova thermometers that is directly applicable to V838 Herculis. Using the present rate in more nova simulations at lower WD masses will provide a range of accurate calibration points, independent of nuclear uncertainties.

Research at NSCL was funded by the National Science Foundation under Grants No. PHY-1913554, No. PHY-1102511, No. PHY-1565546, No. PHY-1811855, No. PHY-2011890, as well as by the Department of Energy Office of Science under Award No. DE-SC0016052. We acknowledge support from the Natural Sciences and Engineering Research Council of Canada (NSERC), as well as the Spanish MINECO grant AYA2017-86274-P, the E. U. FEDER funds, the AGAUR/Generalitat de Catalunya grant SGR-661/2017, and the EU Horizon 2020 grant 101008324 ChETEC-INFRA. This article also benefited from discussions within the ChETEC COST Action (CA16117). Additional funding sources include Korean NRF grant Nos. 2020R1A2C1005981 and 2016R1A5A1013277. We thank the NSCL staff for providing technical and administrative support for this experiment.

\bibliography{LF18052arXiv}

\end{document}